\documentstyle[epsfig,subfigure,float]{article}
\begin{document}

\vspace*{-1.8cm}
\begin{flushright}
{\bf LAL 98-74}\\
October 1998
\end{flushright}
\vskip 7.5 cm{

\begin{center}
{\huge {\bf Search for Higgs Bosons in e$^+$e$^-$ Colliders}}\\

\end{center}

\vskip 1.5 Cm
\begin{center}
{\Large {\bf F. Richard}  \\ }
\end{center}

\vskip 1cm

\begin{center}
{\it \Large \bf Laboratoire de L'Acc\'el\'erateur Lin\'eaire}\\
{\it IN2P3-CNRS et Universit\'e de Paris-Sud, BP 34, F-91898 Orsay Cedex}
\end{center}

\vskip 3 Cm
\begin{center}
{\it Talk given at the Zuoz summer school on Hidden Symmetries and Higgs Phenomena, }\\
{\it Zuoz (Engadin) 16-22 August, 1998}
\end{center}

\vfill\eject
\pagestyle{plain}
\section*{Introduction}
%\vskip 0.5 cm
In spite of theoretical and experimental outstanding achievements, particle
physics has not been able to reach a conclusion on the simple question: from 
which mechanism does mass originate ? It is a common belief that such a
mechanism will be characterized by the observation of at least a scalar
particle. Whether this object is elementary (SM or MSSM scenario), composite
(technicolor scenario) or too heavy to be observed as a particle (cf. M.
Chanowitz'talk at this school) remains uncertain. \par
  In this talk, I will discuss the first scenario and argue that a leptonic
collider provides the best tool to study the properties of the Higgs boson(s).
In the first chapter, I will recall the theoretical and experimental arguments in
favour of this scenario and the predicted properties of the neutral Higgs
boson within the SM scheme and its supersymmetric extensions (for a thorough 
discussion of the theoretical aspects, I refer the reader to the talk given by 
Z. Kunszt at this school). \par
 In the second part of my talk, I will describe the strategy followed at LEP2
for the neutral Higgs boson searches, discuss the present results and future prospects. \par
 In the third part, I will go through the future lepton colliders under study and
give some examples of what could be achieved with such machines. \par 

%\vskip 0.5 cm

\section*{Theory}
%\vskip 0.5 cm

\subsection*{The SM scenario} 
%\vskip 0.5 cm

   In this scenario one simply assumes that there is a single scalar complex
iso-doublet\cite{review} $\Phi$ "which does everything": \par

   - It gives masses to W$^\pm$ and Z bosons (and leaves the photon massless)
through the Higgs mechanism. The masses are given in terms of the vacuum
expectation of $\Phi$, v,
and of the gauge couplings. 3 components of $\Phi$ are "eaten" 
to create the
longitudinal components of W$^{\pm}$ and Z, while the last degree of freedom left is
identified to the Higgs boson.  \par

   - It generates the fermion masses (with the exception of neutrinos) through 
the Yukawa couplings. As a consequence, the Higgs boson couples to the fermions
proportionally to their masses. \par

   This scenario is clearly minimal in the sense that it implies a common source
for all masses. Nature could clearly be less "economical" and use different 
fields for both purposes. As we shall see in describing the detectability 
of the Higgs boson through b-tagging, this would have serious consequences on
the on-going experimental searches.  \par  
   
   The scalar potential of $\Phi$ can be written: 

$$V(\Phi)=\mu^2|\Phi|^2+\lambda|\Phi|^4$$  
    
   The Higgs mechanism, i.e. V minimal for a non zero expectation value of 
$\Phi$, requires 
a negative value for $\mu^2$ ($\lambda$ is the quartic dimensionless coupling which has 
to be positive for V to be bounded from below). The dynamical origin of this
negative mass is not explicit in the SM.  \par
    The mass of the Higgs boson is simply given by m$^2_H$=2$\lambda$v$^2$ where v is
the vacuum expectation of $\Phi$ 
simply related to the W boson mass. While $\lambda$
is not explicitly known, it can be severely constrained if one assumes that the 
SM remains perturbative up to very high energy scales (Planck or unification).
Typically one finds that m$_H<$200 GeV. The requirement of vacuum stability 
imposes a minimum value of about 140 GeV which can be relaxed if new physics 
appears at an energy scale well below the Planck scale\cite{bound}. \par 
 Loop corrections to the mass term being u.v. divergent would give  corrections
of the order of the Planck scale. The SM alone assumption is therefore not 
acceptable (the hierarchy problem). One accordingly assumes that either new 
physics occurs before the TeV scale, as in technicolor\cite{technicolor} or SUSY, or one avoids the
appearance of a  light Higgs boson by postulating a strongly interacting 
sector related to electroweak symmetry breaking (e.g. the so-called BESS
model\cite{bess}).  \par
 Another possibility, recently proposed in \cite{grav1tev}, is to assume that
gravitation becomes a strong force at the TeV scale, therefore providing a natural
u.v. cut-off to the theory. \par
 As a final remark, one should emphasize the incompleteness of the theory in the
fermion sector. There is a complete arbitrariness in the values given to the
Yukawa coupling constants without mentioning the neutrino mass aspects which
fall outside the SM framework. 
%\vskip 0.5 cm

\subsection*{Higgs mass from precision measurement}
%\vskip 0.5 cm

 From previous discussions, it seems that we have no clear theoretical 
guidance on the existence of a light Higgs. Fortunately there is an indirect 
source of information relying on the precise measurements performed at 
LEP,\linebreak
\newpage
\noindent
SLC and FNAL. The SM allows to compute loop corrections\cite{loopz} 
which can be measured precisely
from various independent observations at LEP/SLC: 

$$M_Z,\sigma_{f\bar{f}},\Gamma_{f\bar{f}},A^f_{FB},A^e_{LR},P_{\tau},M_W,
\sigma_{WW}$$
All quantities, at the tree level, are known in terms of the 3 most precise 
quantities: \par
$$M_Z, G_F~~ {\rm and}~~ \alpha.$$  
 The main correction comes from the running of $\alpha$ up to the Z mass and we will come back to
this point shortly.
 The second main correction comes from loops involving the top quark and 
depends
quadratically on the top mass. It can be computed precisely from the 
measurement of the top mass coming from FNAL so that one
can separate the effect
from the Higgs contribution. The loop effect has logarithmic dependence on the Higgs
mass and therefore 
provides a rather imprecise determination of m$_H$. Nevertheless, with significant
improvements on the experimental side (more precise results and internal 
consistency between the various measurements) and on the theoretical 
computations (h.o. terms well under control which means that the calculations 
are meaningful for Higgs masses up to 1 TeV\cite{loop2}) one can safely give an upper 
limit on the Higgs mass in the SM\cite{vancprec}:   

       $$ m_H<280~ {\rm GeV~ at}~ 95\%~ {\rm C.L.}$$ 
        
The corresponding central value is still rather imprecise: 
       $$m_H=84^{+91}_{-51} {\rm GeV}$$   
 
The computation of this result relies on $\alpha$(M$_Z$) which
is known to $\pm$710$^{-4}$ due to our poor knowledge of the e$^+$e$^-$ hadronic
cross-section. Using $\tau$ hadronic decays measurements plus some theoretical
inputs,\cite{davier} has shown that this error can be reduced to $\pm$310$^{-4}$. This
improvement has the effect of consolidating the upper bound on m$_H$ as shown in
figure 1. \par 
In conclusion, one can say that precision measurements put severe restrictions on the mass domain allowed for
the Higgs boson within the SM but it is fair to add that this domain can be extended if new physics appears at finite mass scales
(see e.g. \cite{gatto}). 
\begin{figure}[h]
\centering
%\epsfysize11cm
%\epsfxsize10cm
%\epsffile{Higgs_precis.eps}
\epsfig{figure=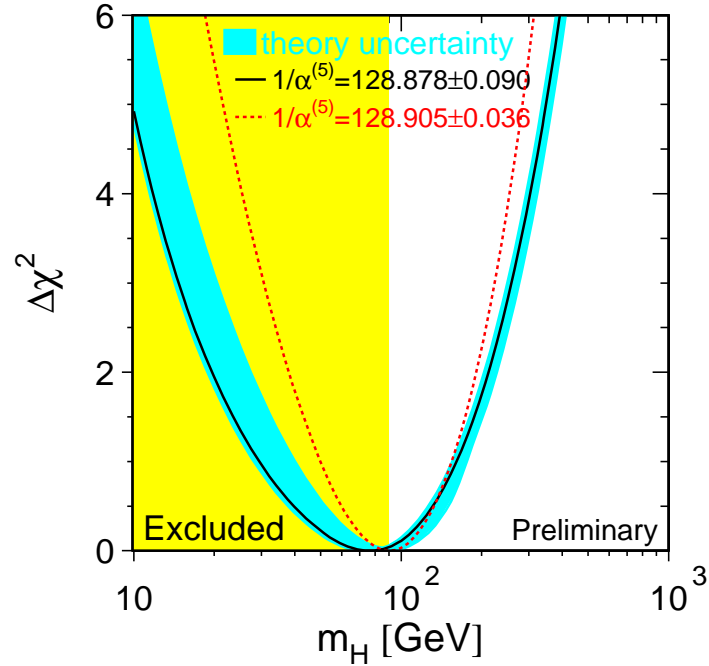,width=9.5cm}
\caption{ $\chi^2$ distribution obtained from the 
precision measurements. The dotted curve corresponds to an 
improved estimate of $\alpha(M_Z)$.}
\end{figure}  
%\vskip 0.5 cm
\newpage
\subsection* {The MSSM scenario} 
%\vskip 0.5 cm
   The minimal SUSY extension of the SM offers a viable solution to the
hierarchy problem provided that the SUSY partners have masses in the TeV region
 or
below. The theory therefore remains finite up to the Planck or GUT scale. A 
simple
picture of the origin of the EWSB effect emerges: $\mu^2$ starts from a 
positive
value at the GUT scale and, under the influence of the top Yukawa constant,
runs to a negative value at the EW scale. 
The $\lambda$ coefficient, and therefore the Higgs mass is simply related
to the gauge couplings. \par 
   The price to pay, in the Higgs sector, is the appearance of two Higgs 
doublets $\Phi_u$ and $\Phi_d$ 
which are required to provide the fermion mass
terms. One can simply justify this by saying that a single Higgs boson model 
is not 
viable within SUSY since it
would have a single fermion SUSY partner which would generate a triangle 
anomaly. \par
    Five physical particles originate from this scheme: h and H the light and
heavy CP even components (which are mixtures of $\Phi_u$ and 
$\Phi_d$ components with a
free mixing angle called $\alpha$), a CP odd component A, two charged 
components 
H$^{\pm}$. Two vacuum expectations occur which can be parametrized as vcos$\beta$ and vsin$\beta$, where 
$\beta$ is an unknown angle. \par

%\vskip 0.5 cm

\subsection*{Expected Higgs mass} 
%\vskip 0.5 cm

\subsubsection*{Within MSSM} 
%\vskip 0.5 cm

     Two parameters are sufficient to describe the Higgs sector. For instance
one has m$^2_h$=F(m$_A$,tan$\beta)\rightarrow$m$^2_Z$cos$^22\beta$ when m$_A\gg$m$_Z$. \par  
     Large loop corrections affect this simple scheme and occur mainly through
the top/stop sector. One can simply write : 

   $$ m^2_h=m^2_Z cos^22\beta+3m^4_t/4\pi^2
v^2 ln(m_{\tilde{t_1}}m_{\tilde{t_2}}/m^2_t)
+\tilde{A^2_t}F(m_{\tilde{t_1}},m_{\tilde{t_2}},\tilde{A_t})$$
where m$_{\tilde{t_1}}$ and m$_{\tilde{t_2}}$ are obtained 
by diagonalizing the 
stop mass matrix given by:
\[ \left( \begin{array}{cc}
m^2_{\tilde{t_L}} & m_t\tilde{A_t} \\
m_t\tilde{A_t} & m^2_{\tilde{t_R}} 
\end{array} \right) \]
where $\tilde{A_t}$ is called the mixing parameter. \par
The second term in the expression giving m$^2_h$ clearly shows the strong influence of the top 
mass in these loop corrections. \par
     In this framework, the upper bound on the Higgs mass, initially m$_Z$, can 
reach up
to 125 GeV when tan$\beta$ is large (i.e. when the 1st term is maximum). \par
    The following question naturally arises: what is the most likely value for
m$_h$ ? \par
    The answer can be given with an assumption on the GUT behaviour of the top
Yukawa coupling Y$_t$. If this value is not too small (here I assume
tan$\beta<$10 such that one can neglect the bottom Yukawa term Y$_b$), 
one observes an infrared fixed point(IRFP) behaviour: whatever
    the value of Y$_t$ at GUT, Y$_t$ converges to the same value at the EW 
    scale. \par
    This gives the following relation: m$_t$=200sin$\beta$ GeV, from which one 
    derives
    tan$\beta\sim$1.6 and therefore\linebreak m$_h<$105 
GeV\cite{casasif}. This upper bound is a 
    safe approximation:
    it includes, for instance, the uncertainty on the calculation due to the
    present error on the top mass.
    There are however objections to this scenario. Firstly, Y$_t$ can be small 
    at
    GUT without creating any problem to the theory. Secondly, there is an other
    IRFP solution tan$\beta\sim$m$_t$/m$_b$, where m$_b$
    is the bottom quark mass. This solution tends to
   create several problems among which a large amount of
   fine-tuning to generate EWSB. This effect can be precisely quantified in the
    following way: given a parameter m from SUSY, fine-tuning is defined as 
    $\Delta$=(dm$^2_Z$/m$^2_Z$)/(dm$^2$/m$^2$). 
    Recently \cite{strumia}
  have argued 
    that, taking into account one loop corrections, fine-tuning is small for 
a large range of m$_h$ values. It is minimal for m$_h$=105 GeV and grows fast 
    above 
    115 GeV when tan$\beta$ becomes larger than 10.  \par
%\vskip 0.5 cm

\subsubsection*{Beyond MSSM} 
%\vskip 0.5 cm
    Since the prediction of a light Higgs boson seems to be an unavoidable
    consequence of the SUSY scheme and, as we will argue later, since this 
    prediction can be tested either with LEP or with future machines, one may ask
    how model dependent is the 125 GeV upper limit obtained within MSSM.  \par
    Beyond the MSSM scheme there is the possibility to introduce an iso-singlet S
    which would have no effect on precision measurements but could help in
    solving the "$\mu$ problem". In the SUSY superpotential one needs a mixing 
    term of the type $\mu\Phi_u\Phi_d$ where $\mu$ is a mass parameter of the order
    of the EW scale which is "put by
    hand". This arbitrariness can be reduced by replacing the previous term by
    $\lambda_1$S$\Phi_u\Phi_d$, and assuming that S acquires a vacuum 
    expectation which generates $\mu$. This term can modify the quartic coupling 
    by $\Delta\lambda$=$\lambda^2_1$sin$^2$2$\beta$ and therefore the resulting 
    Higgs mass. If 
     one requires that the theory remains perturbative up to the GUT scale, one
     can also set a bound on $\lambda_1$ and therefore on m$_h$. This bound 
     increases if one assumes that new physics sets in before GUT.
     Introducing also an iso-triplet field which can 
     couple to the $\Phi_u\Phi_d$ term, the authors of \cite{nmssm}
     achieve an upper bound of 205 GeV. 

%\vskip 0.5 cm
\subsubsection*{Cosmology} 
%\vskip 0.5 cm
  MSSM has the necessary ingredients to produce the right baryon asymmetry in the universe through electroweak baryogenesis.
  This scenario, not unique, requires a Higgs boson lighter than 100 GeV (see \cite{huber} and references therein). \par
    To summarize one has: \par
\begin{description}
\item{-} m$_h<$280 GeV  from precision measurement 
\item{-} m$_h<$205 GeV  from SUSY + new physics before GUT 
\item{-} m$_h<$200 GeV  from SM with no new physics before GUT 
\item{-} m$_h<$125 GeV  from MSSM 
\end{description}         
    This bound can still be decreased by requesting limited
    fine-tuning and/or an IRFP solution (see \cite{gdr} for a more detailed discussion). 
%\vskip 0.5 cm

\subsection*{Higgs production in e$^+$e$^-$ colliders} 
%\vskip 0.5 cm
   In SM, production occurs through the Higgsstrahlung process where a virtual Z* emits a Higgs 
boson and becomes real\cite{review}. This process has a cross-section of a few 0.1 pb provided that 
$\sqrt{s}>m_H+m_Z$. 
There is a sharp energy threshold for Higgs production in
e$^+$e$^-$ colliders and 
one cannot compensate a lack of energy by increasing luminosity as in a
hadron collider. \par
This statement is however partially true since the fusion mechanism, in which two virtual W (or Z) are radiated 
by the incoming leptons and "fuse" into an H, has no sharp threshold.
It turns out however that this process gives very low cross-sections at LEP2 and
only becomes predominant when $\sqrt{s}\gg m_H+m_Z$.     \par
 Other mechanisms, which proceed through loops, like Z$\rightarrow$H$\gamma$\cite{hhg} or Z$\rightarrow$Hgg \cite{hgg},
give negligible contributions unless enhanced by anomalous couplings \cite{hagiw}.\par 
   In MSSM, one has $\sigma_{hZ}=\sigma_{SM}$sin$^2(\alpha-\beta)$. When this process is extinct due
to mixing, a complementary channel, Z$\rightarrow$hA, can be used with  
$\sigma_{hA}=0.5\sigma_{Z^*\rightarrow\nu\bar{\nu}}$cos$^2(\alpha-\beta)\Lambda^{3/2}$, where $\Lambda$ is a
phase-space factor. This process is relevant when m$_h\sim$m$_A$. \par 
   In MSSM one can also produce H either through the Higgsstrahlung process (cos$^2(\alpha-\beta)$ dependence) 
or into HA (sin$^2(\alpha-\beta)$ dependence). 
%\vskip 0.5 cm

\subsection*{Higgs Decays} 
%\vskip 0.5 cm
 In SM, H decays predominantly into bottom quarks when m$_H<$ 130 GeV. Above this mass, couplings to W and
Z pairs become dominant.\par
In MSSM this is also true unless h$\rightarrow$2A becomes kinematically accessible (and usually dominant).
This possibility, disfavoured in usual SUSY schemes which tend to prefer a heavy A, requires very specific searches.
When m$_A>$10GeV, A decays into open beauty and the final state can easily be identified through b-tagging techniques. \par
 A light Higgs boson can decay into gluons (few per cent level) or photons (per mill level) through loops. SUSY contribution, 
mostly from the top squark sector, can affect significantly these modes which are therefore interesting but not
measurable at LEP2.  
A light "fermiophobic" Higgs boson, i.e. without Yukawa couplings, would only decay into photons through a W loop 
and therefore be detectable at LEP2.\par
 As pointed out in the introduction, there is no overwhelming reason to believe that light Higgs bosons necessarily
decay into bottom quarks. Even assuming a scenario where fermion masses 
come from Higgs Yukawa coupling, the
bottom quark coupling in a two-doublet scheme goes like sin$\alpha/$cos$\beta$ and 
can therefore vanish ($\alpha$=0). An alternate scenario, explained in \cite{hhg}, is to assume two-doublets with only one
of them coupling to fermions. This scheme, which falls outside of MSSM, is consistent with FCNC constraints and can lead to
different decay patterns for h and A. \par
 A Higgs boson may also decay invisibly. In the SUSY scenario it can go into neutralino, gravitino or sneutrino\cite{invis}. 
 In models with lepton number violation it could decay into majorons\cite{invisJ}. 
The Higgsstrahlung process, in which the Z 
 can be used as a tag, gives an excellent tool to identify these modes. \par
 In conclusion, Higgs searches should be open minded and take full
 advantage of the clean experimental environment offered by lepton colliders.     
%\vskip 0.5 cm

\section*{Searches at LEP2} 
%\vskip 1. cm

\subsection*{Machine parameters} 
%\vskip 0.5 cm
  In a circular machine with a radius R, synchrotron radiation energy losses grow like $\gamma^4/R^2$. Since it is
impractical to increase further R, LEP can therefore be seen as 
the last high energy circular collider. At 100 GeV,
  electrons loose 3GeV/turn which, for an RF gradient of 6MV/m, requires an accelerating length of about 500 m. With a current of 
 2x4 mA,
 one has to provide 24MW to the beam, and therefore draw about 10 
times more AC power with warm cavities. This figure 
justifies the major technological effort invested by CERN in building
 supraconductive cavities. 272 copper cavities with thin niobium film 
have been produced by european industries under the supervision of
  CERN. This represents a major technical challenge: e.g. these cavities correspond to 2000 m$^2$ of Nb film 
without any defect at the mm$^2$ scale.\par
A center of mass energy of 189 GeV has already been reached
 in 1998 and the plan is to reach 200 GeV in 1999 by raising the average field value to 7 MV/m. Extra cooling power will also be
needed which will require an important upgrade of the cryo-plants during the 1998-1999 shut-down. \par
  The maximum luminosity is now reaching 10$^{32}cm^{-2}sec^{-1}$. In practice the relevant figure is the integrated luminosity  
which, taking into account the efficiency of the machine and the effective 
life-time of the beams, is roughly equal to the maximum
luminosity multiplied by the elapsed time and 
divided by $\sim$2$\pi$. This figure is not unusual with colliders and one should bear 
this in mind when estimating the physics potential of a given machine. In 1997 about 60 pb$^{-1}$ were collected in each experiment
and this figure will increase to about 150 pb$^{-1}$ in 1998.\par
   Taking into account that LEP2 will also run during year 2000, one may reasonably expect that each experiment will accumulate
   a luminosity of 200 pb$^{-1}$ at a center of mass energy of 200 GeV.

%\vskip 0.5 cm

\subsection*{Experimental Tools} 
%\vskip 0.5 cm

   Primordial backgrounds (figure 2) are W pair production, QCD four jet final states q$\bar{\rm q}$gg and ZZ. 
Assuming that the Higgs boson decays into b$\bar{\rm b}$
allows to eliminate the first component(see figure 3) 
and suppresses the two others. 
\par
The q$\bar{\rm q}$ component is dominated by radiative return to the Z resonance. 
This type of effect has a definite pattern and
can be removed. The 4 jet final states (q$\bar{\rm q}$gg) has a cross-section of a few pb before b-tagging. \par
If m$_H\sim$m$_Z$, as is the case in the range of LEP2, 
ZZ appears as
an "incompressible background" since Z bosons decay into b$\bar{\rm b}$ in 
about 17 $\%$ of the cases.
\vskip 2cm 
\begin{figure}[h]
\centering
%\epsfysize14cm
%\epsfxsize14cm
%\epsffile{LEP200_CROSS_SECTIONS.eps}
\epsfig{figure=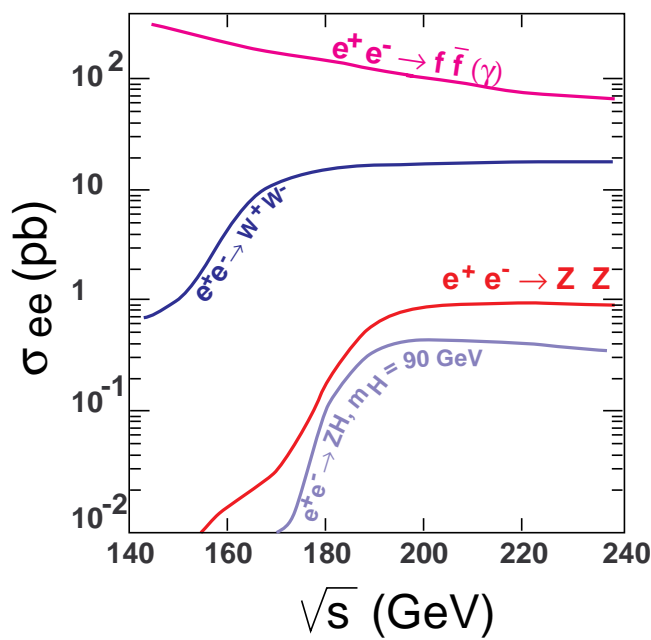,width=12cm}
\vskip -2cm
\caption{Relevant cross-sections for LEP2.}
\end{figure}
B-tagging relies on the measurement of several charged tracks not pointing to the main vertex. Charm decays can be well 
separated since
they give a lower multiplicity. The detector resolution effect is directly measured using the sample of 
non pointing tracks which seem
to originate from a particle decaying upstream of the main vertex. One then constructs a probability function which is uniformly
distributed between 0 and 1 for light quarks. Bottom decays correspond to a very low probability. More involved methods have been 
developed which include the definition of secondary vertices, the mass of the particles at these vertices etc...\cite{boris}\par
   A tremendous effort went into the tuning of this tagging method\cite{borisvan} 
with the motivation of measuring R$_b$, the b fraction of 
Z hadronic decays at LEP1. In view of LEP2, the Si detectors were improved with some emphasis on the solid angle
coverage. \par
   An illustration of the efficiency/rejection of the method is shown in 
figure 3.
\vskip 3cm
\begin{figure}[h]
\centering
%\epsfysize14cm
%\epsfxsize14cm
%\epsffile{DELPHI_BTAG_REJECT.eps}
\epsfig{figure=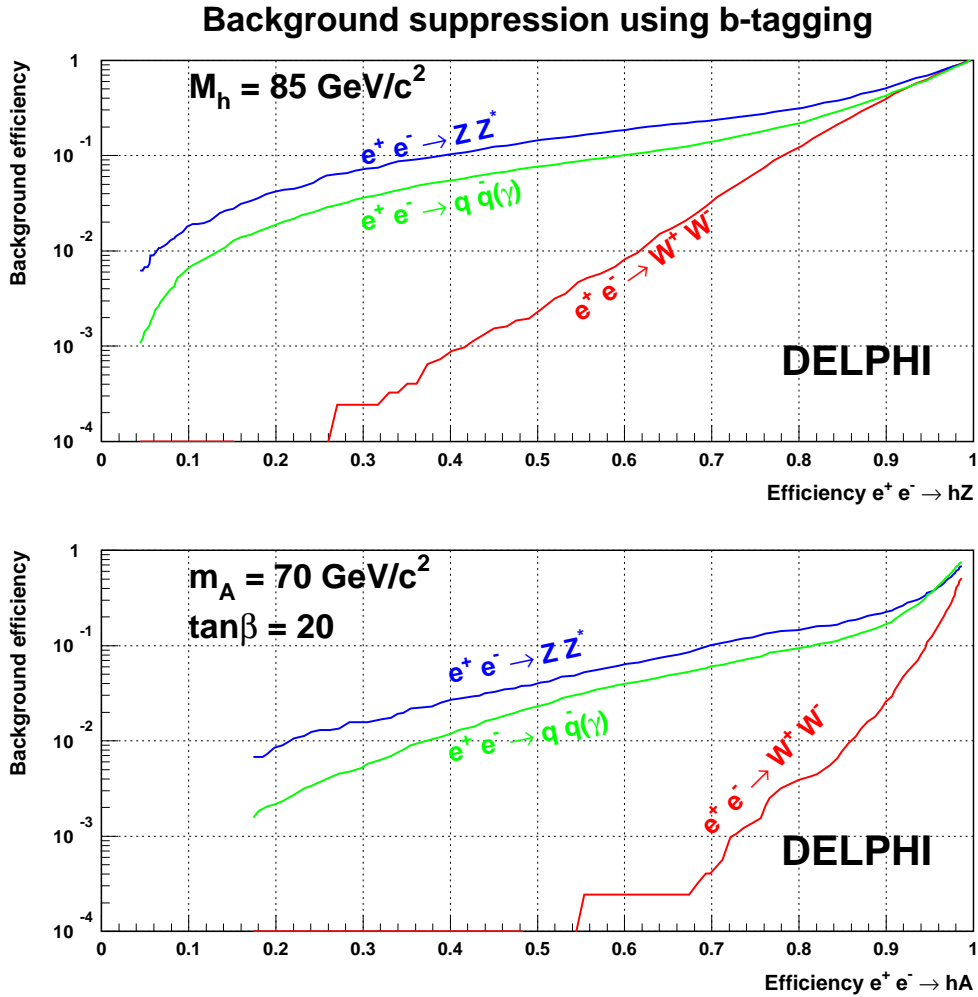,width=14cm}
\vskip -2cm
\caption{Background suppression versus efficiency: top curves are given for h(85GeV)Z, bottom
curves for hA with m$_h\sim$m$_A\sim$70 GeV.}
\end{figure}
%\vskip 0.5 cm

\section*{Results} 
%\vskip 0.5 cm
  The various analyses performed in the 4 LEP experiments include b-tagging plus some extra selections, mostly against
QCD background (the two gluons emitted in 4 jet events tend to be soft and collinear to the quark jets). One can either 
build a global probability which integrates all selective variables or train a neural network to separate the Higgs signal
from the SM processes. 
The key element of this procedure is a careful check of data/simulation agreement at the various levels. This requires 
a good tuning of the SM generators and of the detector response. \par
 An important element of the procedure is an optimal choice of the cut on the selective variable(s), of the combination
of the various channels (e.g. corresponding to the different Z decay modes) and, at the final stage, of the 4 experiments.   
This optimization procedure works automatically on the basis of simulation, avoiding any bias coming from the data. \par
\newpage
Typical figures are given in table I. 
\begin{table}[h]
\centering
\caption{Typical results for LEP2 Higgs searches with 50 pb$^{-1}$}
\vskip 0.5 truecm
\begin{tabular}{|c|c|c|c|c|c|c|}
\hline
  & Eff$\%$ & BR $\%$ & ZZ & QCD & WW & Signal \\
H$\nu\bar{\nu}$  &  30 & 20 & 0.2 & 0.2 & 0.1 & 1.25  \\
\hline H$\ell\ell$       &  60 & 7  & 1   &     &     &  0.7 \\
\hline HJJ       &  40 & 70& 1.3 & 1.6 & 0.8 & 4 \\
\hline
\end{tabular}
\end{table}

It is worth noticing that the figures quoted in this table are far better 
($\sim$twice more efficient in the case of the 4 jet channel)
than anticipated during the workshop\cite{work} preparing
the LEP2 program. \par
Data taken in 1997 have been statistically combined through the Higgs Working Group. For the SM the final result 
is\cite{treille}: 
$$m_H>~89.8~ GeV~ at~ 95\%~ C.L.$$
No excess is observed in the number and the mass distribution of the candidates. The limit obtained is
very close to expectation.\par
The gain obtained by combining the 4 experiments is +3 GeV with respect to the expected limit from 
individual experiments. Clearly one may obtain a similar limit on the basis of one "lucky"
experiment, that is an experiment for which the background has fluctuated negatively yielding
a better mass limit than expected. Although not obviously incorrect, this choice, frequently done,
does not seem the safest. \par 
The hA channel has been searched mainly in b$\bar{\rm b}$b$\bar{\rm b}$ final state for which
the analysis is almost background free and has an efficiency above 50$\%$. \par
Combining hZ and hA searches, one can try to exclude the MSSM scenario in terms of
2 parameters, e.g. m$_h$ and tan$\beta$. Loop corrections introduce additional parameters, like the
squark top masses and the mixing parameter $\tilde{A_t}$. Following the recommendation of the LEP2
workshop\cite{work}, one usually assumes an average squark mass of 1 TeV and $\tilde{A_t}$ either small,
"no mixing case", or such that it maximizes the loop corrections, "large mixing". As
can be seen in figure 4 the effect of these hypotheses reflects in the definition
of the limits of the domain of parameters allowed by the theory. 
One can notice that the present result eliminates the low tan$\beta$ region in the 
no mixing case which corresponds to the IRFP solution. \par
A more correct theoretical treatment\cite{desch}, involving detailed scans of the SUSY parameters has
been performed by the LEP collaborations with the conclusion that results do not change significantly except
for very narrow windows of parameters. \par 
Specific searches were carried out for the Higgs decaying invisibly with the conclusion that the mass limit on m$_h$ is above
80 GeV\cite{treille}. The "fermiophobic" case, with h decaying into 2 photons, is also well treated \cite{desch}. The
general two-doublet model, without assuming b decays, has weak limits \cite{desch}.\par 
One should recall the caveat about the influence of the h$\rightarrow$2A scenario. For a given tan$\beta$,
the minimal value for m$_h$ corresponds to m$_A$=0 and therefore to the lowest limit of the allowed m$_h$ domain in figure 4
(e.g. for tan$\beta$=1 and m$_h\sim$ 55 GeV).
It seems fair to say that no dedicated effort has been devoted to prove that such a possibility is
experimentally excluded. One may however argue that in supergravity and in the gauge mediated SUSY breaking schemes a light A is
excluded \cite{erler}: m$_A>$115 GeV. \par 
  The limit on m$_h$ is weaker at large tan$\beta$. This effect is due to the hA channel. If one again assumes, as suggested by
  theory, that A is heavy, then the limit on m$_h$ becomes independent of 
tan$\beta$. 

\subsection*{Prospects for the future} 
%\vskip 0.5 cm     
 Assuming $\sqrt{s}$=200 GeV and 200 pb$^{-1}$ per experiment, one can expect a
discovery reach up to 107 GeV or an exclusion up to 109 GeV for the SM Higgs\cite{gross}.
The MSSM coverage is shown in figure 5 for the maximal mixing case\cite{janot}. One can see that the IFRP 
scenario is well covered but, given the uncertainty on the top mass which can
move the Higgs maximum mass by a few GeV, it is very important to collect more
than 100 pb$^{-1}$ per experiment. \par
 Again, if one assumes that m$_A$ is heavy, the limit on m$_h$ will not depend on tan$\beta$. For a complete coverage of MSSM, one would need 
m$_A>$(m$_h)_{max}$, a condition not yet reached in \cite{erler}, but which could be reached by the time LEP2 is at its
maximal energy.
 A full coverage of MSSM within supergravity therefore requires $\sqrt{s}\sim$215 GeV, a value tantalizingly close to the potential
of LEP2. It would mean 
either reaching $\sim$9 MV/m or increasing  by 30$\%$ the number of cavities. There are other
limitations discussed in \cite{lepmax}.   
\newpage
\def \textfraction{0}
\begin{figure}[htp]
\centering
%\epsfysize14cm
%\epsfxsize14cm
%\epsffile{LEP_JANOT_TGB_MH.EPS}
\epsfig{figure=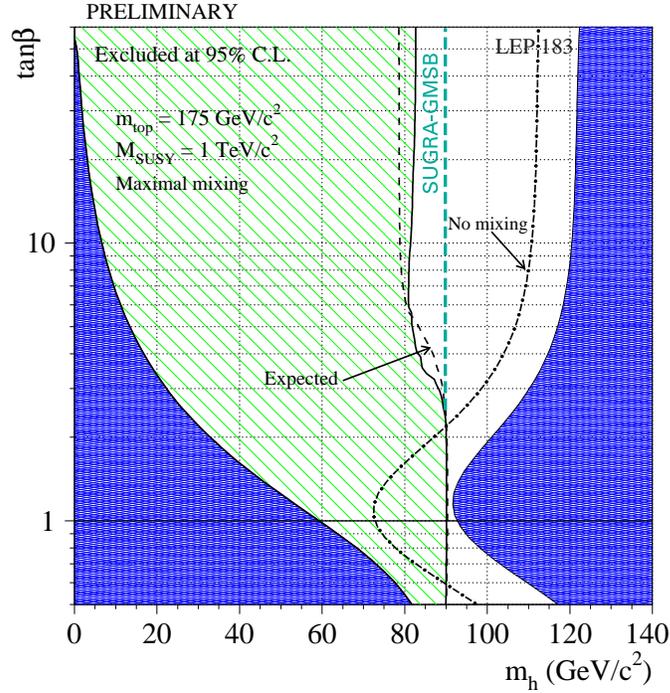,width=10cm}
\caption{MSSM limits reached by combining the 4 LEP experiments (1997 data only). The dashed line
labelled SUGRA-GMSB indicates the limit reached assuming Supergravity or 
the Gauge Mediated SUSY breaking schemes. }
\end{figure}
\vskip 2cm
\begin{figure}[htp]
\centering
%\epsfysize14cm
%\epsfxsize14cm
%\epsffile{JANOT_PROSPECTIVE.eps}
\epsfig{figure=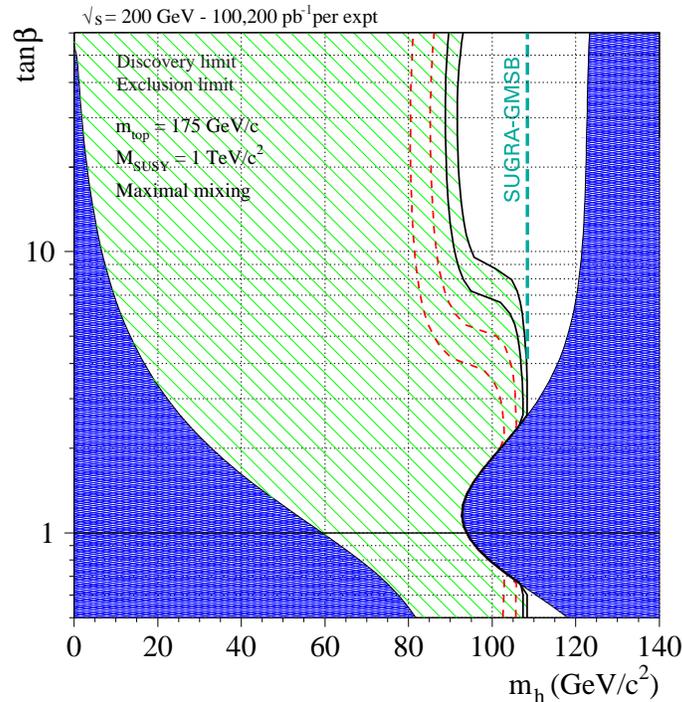,width=9cm}
\vskip -0.5cm
\caption{Expected MSSM coverage at LEP2. The two neighbouring dashed lines correspond to discovery limits reachable
with 100 pb$^{-1}$ and 200 pb$^{-1}$ per LEP experiment at $\sqrt{s}$=200 GeV. The full lines are the corresponding
exclusion limits. The SUGRA-GMSB dashed line is defined as in figure 4.}
\end{figure}
\def \textfraction{0.2}
%\vskip 0.5 cm
\newpage
\section*{Future Colliders\cite{hubner}} 
%\vskip 0.5 cm     
  SM Higgs search at LHC (cf. presentation of D. Denegri), for masses below 130 GeV, use the decay mode into 2 photons. 
This channel
is not always accessible in MSSM, specially at low m$_A$, but can be covered indirectly using complementary channels. It is 
however fair to say that a
full study of the Higgs boson properties can only be done at a lepton 
collider. \par
 Muon colliders\cite{mucol} are very promising in two respects. 
They can reach energies well above 1 TeV and therefore cover an energy reach
 comparable or even beyond LHC. For what concerns the Higgs sector they offer the possibility of single H production with
 reasonable rate given the Higgs 
coupling to muons and the excellent energy resolution of these beams. Before a realistic project can be
launched, a long and active R$\&$D program is needed to test the various 
elements of this complex scheme.

Electron linear colliders have entered in an intensive R$\&$D period to demonstrate the feasibility of the various schemes under
 consideration. One may distinguish 3 approaches: \par

   - The high frequency approach which allows to reach high gradients ($>$50 MV/m) under study at SLAC and KEK. The
     limitations for this scheme are the power sources (several 1000 klystrons are needed) and the wakefields, i.e. fields induced
by the beam bunches which travel closely to RF walls which can distort the beam. It is fair to say that this type of
approach has been progressing very well but seems  limited to 1-2 TeV center of mass energy. \par
    - A two beam accelerator, in which an auxiliary low energy beam generates the power and replaces the many klystrons is
    generally acknowledged as the most promising solution to go beyond 1-2 TeV. CERN has been actively testing this idea and has
    recently re-designed the auxiliary beam into a more realistic solution. This ambitious scheme, known as CLIC, aims at 100 MV/m
    gradient and 30GHz frequency. The luminosity goal is $>$10$^{34}cm^{-2}sec^{-1}$ with a maximum energy of 5 TeV. 
    The limitations 
    of this scheme are the wakefields and a very critical alignment. \par

   - A supra-conductive solution, TESLA, is under active study at DESY. This solution can only work at 
low frequency, low gradients ($<$40
     MV/m). It therefore has a limited energy, below 1 TeV but can provide very high currents, with very good duty cycle, which
     could allow to reach luminosities $\sim$10$^{34}cm^{-2}sec^{-1}$. \par

 Common to all these schemes is the request for a vertical spot size of a nanometer at the interaction point. The FFTB 
experiment\cite{fftb} has
been able to reach a beam spot size of
$\sim$60 nm, but there is still some way to go (e.g. improvement of the emittances in storage rings now
tested at KEK) before one can reliably count on the final figure. \par
     In the parameter list for these projects, one should notice that the energy loss experienced during beam-crossings is 
at the few $\%$ level in such a way that one keeps almost intact the possibility of measuring inclusively (i.e. including the
invisible decay modes) the process HZ  using the leptonic Z decay, plus energy momentum conservation. This allows to
observe a clean and narrow mass peak and deduce the total cross section. \par   

     For what concerns light Higgs physics, energy does not seem an issue 
but luminosity can be of interest, 
as will be discussed shortly. The main uncertainty, in the MSSM case, is on the H and A bosons masses which could be heavy and
therefore require more than 1 TeV center of mass energy for associated production e$^+$e$^-\rightarrow$HA.  

%\vskip 0.5 cm

\subsection*{Higgs Physics at Future Colliders} 
%\vskip 0.5 cm     

  With a luminosity x1000 with respect to LEP2, one could produce several 10$^4$ HZ events, which would provide a clean sample
if m$_H$ is sufficiently distinct from m$_Z$. Precise measurements become possible, including the rare modes into two
photons and two gluons. For the latter, this requires a serious improvement of the tagging techniques to separate this
state from beauty and charm decays. This challenging goal seems reachable given the superior track extrapolations which can
be achieved in a linear collider with reduced beam pipe size (1cm radius instead of 5cm at LEP). First estimates indicate that 
for Z decays it would be possible to tag charm decays with a purity of 
$\sim$80$\%$ and an efficiency of $\sim$60$\%$
\cite{lund}. \par
 The measurement of the ratio 
c$\bar{\rm c}$/b$\bar{\rm b}$, 
which seems experimentally feasible, can be of great interest to distinguish between SM and
 MSSM. Provided that tan$\beta>2$, one can show that this ratio deviates by 
more than 20$\%$ from the SM if m$_A<$450
 GeV\cite{bbcc}.
 Unfortunately this ratio also depends on the effective charm mass at the Higgs mass scale, a quantity poorly known given the large
QCD corrections. The present uncertainty due to this effect is $\sim$15$\%$\cite{cc/bb}. \par
  The Higgs total width can be measured with a muon collider to a fraction of an MeV. For an electron collider, one
can use an indirect method. $\Gamma_{H\rightarrow2\gamma}$ is measurable in an 
$\gamma\gamma$ collider\cite{berk}. This type of collider has been envisaged and it 
was shown that, shooting with a laser on an electron beam, one can generate 
photons which carry 80$\%$ of the incident energy.   
The Higgs total width is deduced from BR(H$\rightarrow2\gamma$) and $\Gamma_{H\rightarrow2\gamma}$. \par
 The Yukawa coupling to fermions is accessible in t$\bar{\rm{t}}$H final states \cite{review}. The cross-section is at the fb level. \par   
  Given the very large luminosity considered for TESLA, one could also 
test the
self coupling H$\rightarrow$2H which simply originates from the quartic term of the Higgs Lagrangian (the $\lambda$ term).
The largest contribution comes from the Higgsstrahlung diagram in which a virtual Higgs H$^*$ is emitted and subsequently decays
into 2 on-shell Higgs bosons. This process gives a very distinct final state HHZ. Unfortunately the cross section, assuming 
m$_H\sim$ 100 GeV, is 0.3 fb at $\sqrt{s}$=500 GeV\cite{hhh} and therefore one needs the highest possible luminosity. \par 
 The heavy Higgs bosons from MSSM, H$^{\pm}$, H and A, could be pair-produced up to masses of $\sqrt{s}$/2.
Single production of H and A is possible at a muon collider with excellent mass resolution, a key feature 
since H and A tend to be degenerate
in mass in MSSM. 
   If a $\gamma \gamma$ collider can be operated, it could also allow single production of H and A and therefore increase the
mass range for these bosons up to about 80$\%$ of the maximum center of mass energy. This idea deserves detailed investigation,
in particular concerning the detectability of these channels. \par
   
%\vskip 0.5 cm

\section*{Final Remarks} 
%\vskip 0.5 cm     
       
   After this presentation one may wonder if there could be an escape for a light Higgs discovery in the next 10-15 years ?
\par
   This possibility is clearly present at LHC, even in the MSSM scheme. For instance, as recently shown \cite{topmix}, 
one can concoct a scenario
 with large mixing in the top squark sector in which there is strong cancellation between stop and top loop contributions in the
gluon-gluon fusion mechanism which provides the largest part of the production cross-section at LHC. \par
  Beyond MSSM, there could even be difficult cases for a lepton collider. We have already seen how it was possible to raise 
the upper
bound on the Higgs mass by introducing singlets and triplets. This game has also been played in \cite{gunion}
 assuming a large multiplicity of
Higgs n-plets, including the possibility of invisible decays. The conclusion was that detectability cannot fail given the 
well constrained final states and provided that one can reach an integrated luminosity of 500fb$^{-1}$.  \par

%\pagebreak
%\noindent
\subsection*{Acknowledgements}
%\vskip 0.5cm
 It has been a pleasure to attend this meeting, surrounded by the magnificent landscape of Engadin, and I
warmly thank the organizers for their kind invitation. \par 
 Useful discussions during the preparation of this write up with C. M. Rivero and D. Treille 
are gratefully acknowledged.

\end{document}